# Complex plasmon-exciton dynamics revealed through quantum dot light emission in a nanocavity

Satyendra Nath Gupta[1,&], Ora Bitton[2,&], Tomas Neuman[3,4,&], Ruben Esteban[3,4], Lev Chuntonov[5], Javier Aizpurua[3,4,*], and Gilad Haran[1,*]

[1]Department of Chemical and Biological Physics, Weizmann Institute of Science, POB 26, Rehovot 7610001, Israel,

[2]Department of Chemical Research Support, Weizmann Institute of Science, POB 26, Rehovot 7610001, Israel,

[3]Materials Physics Center CSIC-UPV/EHU, Paseo Manuel de Lardizabal 5, 20018 Donostia-San Sebastián, Spain

[4]Donostia International Physics Center DIPC, Paseo Manuel de Lardizabal 4, 20018 Donostia-San Sebastián, Spain

[5]Schulich Faculty of Chemistry, Technion-Israel Institute of Technology, Haifa, Israel.

*Correspondence: gilad.haran@weizmann.ac.il, aizpurua@ehu.eus

&These authors contributed equally to the work.

**Abstract**

Plasmonic cavities can confine electromagnetic radiation to deep sub-wavelength regimes. This facilitates strong coupling phenomena to be observed at the limit of individual quantum emitters. Here we report an extensive set of measurements of plasmonic cavities hosting one to a few semiconductor quantum dots. Scattering spectra show Rabi splitting, demonstrating that these devices are close to the strong coupling regime. Using Hanbury Brown and Twiss interferometry, we observe non-classical emission, allowing us to directly determine the number of emitters in each device. Surprising features in photoluminescence spectra point to the contribution of multiple excited states. Using model simulations based on an extended Jaynes Cummings Hamiltonian, we find that the involvement of a dark state of the quantum dots explains the experimental findings. The coupling of quantum emitters to plasmonic cavities thus exposes complex relaxation pathways and emerges as an unconventional means to control dynamics of quantum states.

**Introduction**

Manipulating and controlling the interaction of photons with individual quantum emitters has been a major goal of quantum photonics in recent years [1-3]. Such control can be realized by engineering the local photonic environment of the quantum emitter, e.g. by placing it inside an optical cavity [4]. By coupling the excited state(s) of the emitter to the electromagnetic (EM) field of the cavity, one can generate various exotic light-matter coupled states [1,2], single-photon emission sources [5,6], and photonic switches [7,8]. In recent years it has been shown that the formation of new hybrid light-matter states (polaritons) within optical cavities can dramatically affect photophysics [9-11] and chemical reactivity [12,13].

Plasmonic cavities (PCs) formed by metallic surfaces can tightly confine light to deep sub-wavelength regimes [14,15]. The ability to strongly couple quantum emitters to individual PCs has aroused much excitement in recent years [16-19]. Our lab [20,21] and others' [22-25] have demonstrated that such a strong coupling can be realized even in the limit of a single semiconductor nanocrystal (quantum dot, QD) or molecule, and can be observed as vacuum Rabi splitting in light scattering, photoluminescence (PL) or electron energy loss spectra of the coupled systems. Plasmonic cavities with coupled quantum emitters may serve as new testbeds for studies of quantum optical and chemical dynamics under ambient conditions.

In this work, we expose the remarkable modulation of the excited-state dynamics of QDs embedded within PCs through the comparative analysis of an extensive set of light scattering and PL measurements. Using Hanbury Brown and Twiss (HBT) interferometry, we observe non-classical emission from one to three QDs within our devices, and find surprisingly long excited-state relaxation times. We interpret the experimental results using simulations based on an extended Jaynes-Cummings model, which considers the coexistence of bright and dark excited states in the QD with very different coupling properties. We demonstrate the crucial role that the dark state of the QDs plays in the observed dynamics, and thus in shaping of the final PL spectrum.

## Results

**Scattering and Photoluminescence spectra of individual coupled plasmonic devices.**
We used electron-beam lithography to fabricate silver bowties on 18 nm $SiO_2$ membranes. CdSe/ZnS quantum dots (QDs, obtained from MK Impex Corp. with a size of 6-8 nm) were positioned into the gap region of bowties using interfacial capillary forces (Figure 1a). Coupling rates in such devices can exceed 100 meV, depending on the position within the cavity [20]. Scattering spectra of individual QD-bowtie hybrids were measured using dark-field (DF) microspectrometry [20,26], while PL spectra were measured from the same devices following excitation with a CW laser at 532 nm.

Scattering and PL spectra were recorded from 23 bowtie cavities loaded with either one or a few QDs, and two examples are shown in Figure 1b-e (see additional spectra in Supplementary Figure 1). Scattering spectra show dips indicative of plasmon-exciton coupling [20] (Figure 1c & e, green lines). The splitting values obtained directly from the scattering spectra in Figure 1 (i.e. the differences between the two peak positions) are 200 and 290 meV, respectively. Fits of the scattering spectra to a coupled-oscillator model [27,28], presented in Supplementary Figure 2, provide values for the coupling rate, $g$, of 52.6±0.3 meV and 103.5±1.1 meV, respectively. A histogram of the splitting values of all devices is shown in Figure 2a, and a histogram of $g$ values obtained from coupled oscillator fits is shown in Supplementary Figure 3. These values vary between 52.5 meV and 110 meV, with an average value of 71.7 meV.

It is instructive to ask where the coupling rates observed here places our PC-QD systems with respect to the strong coupling regime. To that end, we compare our measured $g$ values to two criteria often discussed in the literature [29]. The first criterion, $g>(\gamma_p-\gamma_e)/4$ (where $g$, $\gamma_p$ and $\gamma_e$ are the coupling strength, plasmon linewidth and exciton linewidth, respectively), guarantees two real solutions in the coupled-oscillator model when the QD is resonantly tuned to the plasmon and can be interpreted as the definition of a lower bound for the strong-coupling regime. When this criterion is fulfilled the system has passed an exceptional point [30,31] and is therefore guaranteed to possess two distinct eigenstates. The splitting in the spectrum above the exceptional point reflects the formation of two polaritonic states. Based on the values of $\gamma_p$ and $\gamma_e$ we measure on bare bowties and free QDs, respectively, the above criterion gives a threshold value of ~55 meV, and the vast majority of the values of $g$ extracted from our spectra are larger.

While this criterion guarantees the existence of two eigenstates, it does not ensure that the measured spectra will clearly show these states as separate peaks. Therefore, a second and stricter criterion is often introduced. This criterion [29], given by $g>(\gamma_p+\gamma_e)/4$, is more heuristic and is connected with the establishment of Rabi oscillations in the time domain. In our case the latter criterion gives a value of ~120 meV, somewhat larger than the values we report here. However, as g increases from the limit given by the first criterion, splitting of the two modes in the spectrum grows continuously, and indeed splitting is systematically present in our experimental spectra. Therefore, we can safely state that the PC-QD systems measured here are found to be at the onset of the strong-coupling regime.

A peak splitting is also observed in PL spectra (Figure 1c & e, red) recorded from the same cavities. PL spectra look significantly and consistently narrower than the corresponding scattering spectra, suggesting that different microscopic mechanisms account for splitting in the two cases. This difference is also manifested in the values of the splitting between peaks obtained from the PL spectra of Figure 1c & e, which are only 100 meV and 110 meV, respectively. The histogram of the peak splitting values obtained from PL spectra is shown in Figure 2b. Comparing this histogram to the one in Figure 2a, we find that while in DF scattering spectra splitting values ($\Omega_{DF}$) are as high as 350 meV, the maximal splitting ($\Omega_{PL}$) observed in PL spectra is only 160 meV. A correlation plot of $\Omega_{PL}$ versus $\Omega_{DF}$ is shown in Figure 2c. It is evident that the correlation is very weak, suggesting that $\Omega_{PL}$ does not depend on parameters like bowtie gap size, number of QDs and others to the same extent as $\Omega_{DF}$.

**HBT interferometry.** To shed light on the observation that $\Omega_{PL}$~const <$\Omega_{DF}$ and further understand the quantum properties of the light emitted by the coupled devices, we turned to HBT interferometry. We first measured the second-order photon correlation curves ($g^{(2)}(t)$) of light emitted from individual QDs on a glass substrate. An example of such a correlation curve is shown in Figure 3a. (Additional examples are presented in Supplementary Figure 4.) The antibunching observed in the correlation curve at zero delay, with a value lower than 0.5, verifies that the measurement is indeed from a single QD. However, in some cases the number of QDs within the laser spot during the HBT measurement was larger than one. In order to obtain both the lifetime of the emitting exciton and the number of QDs, we therefore fitted the measured correlation curves with Eq. (1),

$$g^{(2)}(t) = A + B\left(1 - e^{-\left|\frac{t}{\tau}\right|}\right). \qquad (1)$$

In this equation, A and B are constants and τ is the lifetime of the emitting exciton (i.e. the total decay time). The value of the second-order photon correlation curves at zero time delay, $g^{(2)}(0)$, scales as 1-1/$N$, where N is the number of QDs [32]. However, background photons reduce slightly the zero time dip, which is given by the constant A obtained from the fit. The maximal possible $N$ based on a particular $g^{(2)}(0)$ measurement is the largest integer smaller than 1/(1-$A$). We obtained 22 correlation curves of individual QDs on glass, and used these in order to plot the distribution of exciton lifetimes, which arises due to their non-uniform size distribution (Supplementary Figure 4). The average exciton lifetime is 24 ns, and the distribution is asymmetric with a standard deviation of 5.3 ns.

We then measured the second-order photon correlation curves from QDs within PCs. Two examples are shown in Figure 3b & c. As in the case of QDs on glass, these correlation curves show clear evidence of antibunching, pointing to the non-classical nature of the emitted light. Fitting the correlation curves to eq. 1, we found that the probed devices contained one QD (Figure 3b) and three QDs (Figure 3c). The fits also provided the polariton lifetimes for the two devices: 5.6±0.3 ns and 3.5±0.2 ns, respectively. Overall, $g^{(2)}(t)$ functions were measured from 16 of the devices whose scattering and PL spectra showed a clear indication of peak splitting. More examples of $g^{(2)}(t)$ are provided in Supplementary Figure 5. Supplementary Figure 5d shows the distribution of lifetimes obtained from fits to the correlation curves, ranging from 3 ns to 12 ns. Surprisingly, there seems to be only a minor shortening of the lifetimes (by a factor of ~5) compared to QDs on glass. To verify this result, we also performed direct time-resolved PL measurements of several devices, the results of which are shown in Supplementary Figure 6. The average lifetimes extracted from these measurements are also shortened by a factor of ~5 only compared to the lifetimes of bare QDs (See Supplementary Table 1). This finding is highly unexpected, as the mixing of the QD exciton with the plasmon in the cavity should have opened a fast relaxation channel with a lifetime closer to that of the plasmon [33]. A recent study of an ensemble of QDs deposited on a plasmonic hole array also reported only a modest shortening of the excited-state lifetime [34]. Interestingly, Ebbesen and coworkers found a similar deviation from the expected shortening of the PL lifetime in a different system consisting of molecules coupled to a microcavity [9].

**Extended Jaynes-Cummings model simulations: beyond two levels.** Three surprising observations emerge from the experiments reported above. First, PL spectral peaks of QDs coupled to the PCs are narrower than those in scattering spectra. Second, the splitting between peaks observed in PL spectra is not correlated with the splitting in scattering spectra. Finally, the PL lifetime seems to be only mildly shortened compared to that of QDs on glass. All three observations deviate from expectations for strongly or close-to-strongly coupled QD-plasmonic devices. Indeed, the formation of polaritonic states due to coupling, as described within a standard Jaynes-Cummings Hamiltonian of a two-level system coupled to an optical cavity, would lead to (i) broad scattering and PL spectral peaks, (ii) correlation between the values of splitting seen in the two spectra , and (iii) PL lifetimes on the femtosecond time scale, close to the ultrafast decay times of the PCs. Therefore, the spectral features and the decay of the PL found here indicate a picture that is significantly more complex than that described with a simple coupled two-level model. Indeed, the presence of long-lived dark excitonic states of different origins in QDs has been reported [35-37]. We thus consider the role of a dark state of the QDs as a key contributor to the observed excited-state dynamics of the coupled system.

To simulate such dynamics and examine its potential effect on the experimental observations, we adopt a cavity-quantum electrodynamics (c-QED) theoretical framework. We extend the Jaynes-Cummings Hamiltonian beyond the standard two-level description and include a weakly coupled dark state. We also add Lindblad terms to the Hamiltonian in order to describe incoherent pumping (for the PL spectra) and all the corresponding relaxation channels. The quantum emitter is therefore modeled as an electronic system composed of three levels: a ground state, a level with a large decay rate, representing the lowest bright excitonic state, and another level, positioned slightly lower in energy and possessing a much smaller (yet non-zero) decay rate, representing the dark state. A scheme of the plasmon and quantum emitter energy levels used in this model is shown in Figure 4a, and the relevant parameters (selected to properly describe the physical properties of the system) are given in Table 1. From dynamic simulations based on this model, we calculate scattering and PL spectra, as well as second-order photon correlation functions, which are shown in Fig. 4b and c for a representative case. The relative lack of dependence of of these spectra and correlations on the value selected for the intrinsic decay rate of the dark exciton is demonstrated in Supplementary Figure 7.

Importantly, the simulated second-order photon correlation curves (Figures 4c and Supplementary Figure 8) involve two distinct decay components, a very fast one on the femtosecond time scale, and a much slower one, on the nanosecond time scale. The fast decay component is related to the dynamics of the bright exciton, modified by the involvement of the plasmonic decay channels. However, the time scale of this component is too short to be observed in our experiments. Hence, only the slow decay component of the correlation curve is registered experimentally. This component can be attributed to the decay of the dark state into two possible channels. The first decay channel is due to the population transfer to the bright excitonic mode of the QD, from which fast emission brings the system back to the ground state. The second decay channel involves the enhanced emission due to weak coupling of the dark state to the plasmonic mode. The effect of the coupling of the bright and dark excitons to the plasmon on the resolution of the fast component of g$^{(2)}$ is illustrated in Supplementary Figure 9.

The calculated PL spectra are significantly narrower than the scattering spectra (Fig. 4b), and show a reduced splitting between their emission peaks, in qualitative agreement with the experimental observations. To understand the origin of the two peaks in the PL spectra and to clarify how the bright exciton influences the direct emission from the dark exciton, we perform a series of calculations in which some of the coupling channels are cancelled. We first numerically calculate the photoluminescence spectra for two tailored scenarios in which either the dark exciton or the bright exciton is decoupled from the plasmon. Importantly, while in each scenario only one of the excitons couples to the plasmon, both excitons are still allowed to incoherently exchange populations via incoherent decay and pumping processes. We show in Fig. 5 (a) the PL spectrum calculated with a decoupled dark exciton, $S_{\text{em}}^{g_\text{D}=0}(\omega)$ (green line) and with a decoupled bright exciton, $S_{\text{em}}^{g_\text{B}=0}(\omega)$ (blue line). $S_{\text{em}}^{g_\text{B}=0}(\omega)$ features a single peak due to emission from the dark exciton. Remarkably, $S_{\text{em}}^{g_\text{D}=0}(\omega)$ also shows a single (asymmetric) peak, a signature of the bright-exciton emission at the onset of the strong-coupling regime [23]. This peak is broader than the original exciton peak due to the interaction with the cavity plasmon. As the system is at the onset of strong coupling, the spectral feature appears as an asymmetric peak due to the influence of the not-fully developed upper polariton. If we directly sum the two contributions ($S_{\text{em}}^{g_\text{D}=0}(\omega) + S_{\text{em}}^{g_\text{B}=0}(\omega)$, red dotted line), we can observe that the dark-exciton peak becomes a prominent sharp feature on top of the broader bright-exciton peak. This result strongly differs from the calculation based on the full model, where all the couplings between QD and PC states are considered (shown in Fig. 4 (b), red line). This indicates that the simple sum of uncoupled emissions, $S_{\text{em}}^{g_\text{D}=0}(\omega) + S_{\text{em}}^{g_\text{B}=0}(\omega)$, cannot fully capture the underlying physics of the emission.

We next show in Fig. 5(b) an analytical decomposition of the total PL spectrum, into the contribution from the bright-exciton, $S_{\text{em(B)}}(\omega)$ (green line), and that from the dark-exciton, $S_{\text{em(D)}}(\omega)$ (blue line), while maintaining the couplings $g_B$ and $g_D$ (see details of the model and explicit expressions of this decomposition in Supplementary Note). The total spectrum given by the sum of both contributions, $S_{\text{em}}(\omega) = S_{\text{em(B)}}(\omega) + S_{\text{em(D)}}(\omega)$ (red solid line), agrees very well with the numerical result shown in Fig. 4(b) (see a direct comparison in Supplementary Figure 10).

The contribution of the dark exciton, $S_{\text{em(D)}}(\omega)$, is strongly affected by the bright-exciton coupling. When this coupling is switched on [blue line in Fig. 5b)], the emission of the dark exciton is dramatically reduced compared to the $g_B = 0$ case [blue line in Fig. 5 (a)]. This is due to the formation of hybrid states involving the bright exciton and the plasmon, and to the resulting modulation of the photonic density of states (PDOS), which leads to a lower probability of light emission from the dark exciton and thus reduces the intensity of the dark-exciton contribution. On the other hand, we observe that the emission of the bright exciton is not affected by the coupling of the dark exciton to the cavity [compare green lines for $S_{\text{em}}^{g_D=0}(\omega)$ in Fig. 5(a) and $S_{\text{em(B)}}(\omega)$ in Fig. 5(b)], since the dark exciton only weakly perturbs the plasmonic response. Nevertheless, the presence of the dark exciton still indirectly affects the strength of the bright plasmon, due to incoherent and pumping processes intrinsic to the emitter (included in both models).

We have thus shown that the bright exciton and the dark exciton contribute to the PL spectra in two distinct ways. The complex dynamics of the bright exciton, non-perturbatively coupled to the plasmonic cavity, leads to the formation of an asymmetric emission peak whose line shape deviates from a simple Lorentzian profile. The amplitude of this peak depends on the incoherent pumping of the bright exciton by the external illumination, either directly or via the dark exciton. On the other hand, the dark exciton couples to the plasmonic cavity only weakly, and contributes to the light emission indirectly, via incoherent pumping of the bright exciton, and directly via weak coupling with the plasmonic cavity. Due to the weak perturbative character of the dark-exciton coupling, its direct contribution to the light emission, properly modified by the bright state, can be superimposed on top of the photoluminescence spectrum of the bright exciton (also modified by the dark state pumping). This theoretical model thus shows that the emission probability of the dark exciton via the plasmon is strongly modified due to the modulation of the PDOS by the coupling of the plasmon and the bright exciton. Our results thus indicate that it is essential to consider all coupling and decay channels for a full description of the dynamics and emission properties of the system.

**Discussion**

We reported here a vast set of measurements of QDs embedded within PCs, which allowed us to expose unique excited-state dynamics involving complex interaction between bright and dark states. Coupling values of 50-110 meV were deduced from light-scattering spectra by fitting to a coupled-oscillator model, indicating that our devices are close to the strong coupling limit. In addition to scattering, we also obtained the PL spectrum of each device, and for most devices we also recorded the second-order photon correlation curves. The observation of antibunching in correlation curves demonstrated unequivocally the non-classical nature of the light emitted by these devices, which originated from either one or just a few (countable) QDs. Our PL measurements revealed several deviations from expectations based on the familiar picture of a simple two-level quantum emitter coupled to a cavity resonance. An extended Jaynes-Cummings model that explicitly took into account the presence of a dark state in the QD within a c-QED framework nicely accounts for all the intriguing features in the experiments. This quantum model allowed us to show how the interplay of the weak effect of the PC on the dark exciton and the much stronger effect on the bright exciton leads to the complex dynamics exposed in our experiments.

Theoretical and experimental studies of QDs have revealed different types of dark states. Exchange interactions lead to the splitting of the band-edge exciton with the appearance of a dark state as the lowest energy level and a bright state above it [35]. Experimental work provided direct evidence for this splitting and showed that the dark and bright states are separated by less than 1 meV [38]. This energy difference is too small to account for our observations. On the other hand, the occurrence of trapped surface states whose transitions are significantly red shifted compared to the bright exciton [36,37] can account for the hierarchy of energies used in our model. A location of a few tens of meV to the red side of the PL peak for the surface trap state has already been discussed in the literature, e.g. by Morello et al. [39]. Bradley and coworkers also modeled their experimental data with a dark trap state, though they assign a smaller value of 4-7 meV to the shift of the trap state from the main luminescent state [40]. This value is likely dependent on the particular type of QDs studied, and might in reality be distributed over a certain range. We are not aware of any additional excited states of QDs whose involvement might explain our results. A high-energy shoulder on the PL spectrum of a coupled QD was reported in Ref. 23, and was proposed to be due to a charged exciton or multiexcitonic states. While this high-energy shoulder is likely due to a different origin than the low-energy peak in our spectra, one cannot completely discard the possibility that a similar mechanism contributes to the current results. Nevertheless, our theoretical simulations support the assertion that the low-energy narrow emission line is related to a dark state; the interaction of this dark state with the plasmonic cavity dramatically enhances its emission. This enhancement is still influenced by the plasmon coupling to the bright state, which modulates the final exact contribution of the dark state to the total photoluminescence.

Our findings, based on joint experimental and theoretical observations, demonstrate unexpectedly rich excited-state dynamics induced by coupling of a small number of quantum emitters to a PC. The ability to access and eventually control this complex dynamics of excited states in light emitters can pave the way for manipulation of electronic excitations at room temperature in strongly coupled devices. This is a necessary first step for future applications, such as the construction of quantum devices operating under ambient conditions and the modulation of chemical reactivity at the single-molecule level.

## Methods

**Fabrication of silver bowties.** SiN grids (TEM windows) were cleaned with plasma (O$_2$~ 3.5 sccm and Ar~1.5 sccm) at 150 W. The cleaned grids were spin-coated with polymethylmethacrylate (PMMA) at 4000 r.p.m for 45 seconds to get a 60 nm thick layer of the polyemer, followed by baking at 180 C for 90 seconds. The PMMA coated grids were then transferred to a Raith E_line Plus electron beam lithography chamber for electron beam exposure of PMMA in a series of pre-defined bowtie shapes, using an accelerating voltage of 30 kV and a current of 30 pA. The overall design of each fabricated grid involved matrices of bowties that were separated by 10 μm from each other to avoid any potential interaction between them. Each bowtie was composed of two 80 nm equilateral triangles, so that its plasmon resonance overlapped with the QD emission frequency (see Supplementary Figure 11 for the scattering and PL spectra of an empty bowtie and a QD).The exposed PMMA was developed in a solution containing methyl isobutyl ketone and isopropyl (IPA) alcohol in 1:3 ratio for 30 seconds, followed by dipping in isopropyl alcohol (stopper) for 30 seconds and drying in a N$_2$ gas flow. Subsequently, 3 nm chromium was deposited as an adhesion layer, which was then followed by evaporation of a 20 nm silver layer within an electron-beam evaporator (Odem Scientific applications). Following metal deposition, a liftoff process was carried out using a REMOVER PG solvent stripper to obtain a set of silver bowties on the SiN grid.

**Incorporation of QDs into the gap regions of bowties.** The resist ZEP (a 1:1 copolymer of α-chloromethacrylate and α-methylstyrene) was spin-coated on the bowtie sample at 3000 r.p.m. for 45 seconds, and the sample was then baked for 180 seconds at 180 C. By using alignment marks, the electron beam was positioned at the bowtie gaps with an overlay accuracy of few nm to generate holes in the resist. The exposed regions were developed in amyl acetate and isopropanol. In order to drive QDs into the holes, we followed a method developed by Alivisatos and colleagues [41]. The sample was placed vertically in an aqueous solution of QDs, and the solvent was allowed to evaporate slowly, exerting a capillary force along the receding line of contact, which drove the QDs into the holes. The number of QDs in the gap region could be partly controlled to be one, two or many by tuning the concentration of QD solution and diameter of the holes. A schematic of the bowtie fabrication and QD trapping process is shown in Figure 1a

**Dark-field and PL microspectrometry.** Scattering spectra were measured using a home-built setup based on an inverted microscope and equipped with a 75 W Xenon lamp (Olympus), a dark-field condenser, a 100x oil immersion objective of a tunable numerical aperture (from 0.9 to 1.3), a 150 mm spectrograph (SpectraPro-150, Acton) and an air-cooled CCD camera (Newton, Andor Technologies). A NA of 0.9 was typically used in these experiments. Photoluminescence measurements were performed on the same setup using a NA of 1.3. The excitation source was a 532 nm laser, whose polarization was selected to be parallel to the long axes of the bowties. All the spectra were smoothed with a Savitzky-Golay filter.

**Time-resolved PL measurements.** HBT interferometry and PL decay measurements were carried out using a Micro Time 200 (PicoQuant) single-molecule spectrometer. A 485 nm CW laser was used to excite the QDs through a 60X water-immersion objective. For thes second-order correlation measurements emitted light from the sample was collected with the same objective and passed through a 50/50 beam splitter before being focused on two single-photon avalanche photodiodes (PerkinElmer, 50 ps time resolution). A band-pass filter was inserted in front of each detector to reduce the unwanted background signal. A single detector was used for time-resolved single-photon counting measurements with the same system. In both types of measurement we used a HydraHarp 400 time-interval analyzer (PicoQuant) for signal registration.

**Scanning transmission electron microscopy (STEM).** Plasmonic bowties and QDs were imaged using a Zeiss Gemini SEM microscope in a STEM mode, with an electron beam energy of 30 keV, a 20 μm aperture and a 5 mm working distance.

**Theoretical model.** We assume that a quantum dot is composed of a ground state $|g\rangle$ of energy $E_g$=0 eV, a bright excitonic level, $|e_B\rangle$, of energy $E_B = E_g + \hbar\omega_B$, and a dark level, $|e_D\rangle$, of energy $E_D = E_g + \hbar\omega_D$. The Hamiltonian describing the QD, $H_{QD}$, can be expressed as:

$$H_{QD} = \hbar\omega_B |e_B\rangle\langle e_B| + \hbar\omega_D |e_D\rangle\langle e_D| \quad (2)$$

The QD is coupled to a single mode of a plasmonic cavity of energy $\hbar\omega_{pl}$, described via the Hamiltonian $H_{pl}$:

$$H_{pl} = \hbar\omega_{pl} a^\dagger a, \quad (3)$$

where $a$ ($a^\dagger$) is a bosonic annihilation (creation) operator. The plasmon-exciton coupling is included via the Jaynes-Cummings coupling terms:

$$H_{\text{pl-QD}} = \hbar g_B(a^\dagger |g\rangle\langle e_B| + a|e_B\rangle\langle g|) + \hbar g_D(a^\dagger |g\rangle\langle e_D| + a|e_D\rangle\langle g|), \tag{4}$$

where $g_B$ ($g_D$) is the Jaynes-Cummings constant coupling the bright (dark) level to the plasmon.

The total Hamiltonian, $H$, of the system thus becomes

$$H = H_{\text{QD}} + H_{\text{pl}} + H_{\text{pl-QD}}. \tag{5}$$

To obtain the observables of the system, we solve the Liouville-von Neumann equation for the system's density matrix, $\rho$:

$$\frac{d}{dt}\rho = -\frac{i}{\hbar}[H,\rho] + \sum_i \gamma_{O_i}\mathcal{L}_{O_i}[\rho]. \tag{6}$$

which includes incoherent Lindblad operators, added to account for losses and pure dephasing. These operators take the following form:

$$\gamma_{O_i}\mathcal{L}_{O_i}[\rho] = \frac{\gamma_{O_i}}{2}(2 O_i \rho O_i^\dagger - \{O_i^\dagger O_i, \rho\}), \tag{7}$$

where $O_i$ is a generic system operator to be specified, and † stands for Hermitean conjugate. In particular, we add the following Lindblad terms:

$$\kappa \mathcal{L}_a[\rho] \text{ (Plasmonic decay)}, \tag{8}$$

$$\gamma_{gB}\mathcal{L}_{|g\rangle\langle e_B|}[\rho] \text{ (Decay of the bright excitonic level)}, \tag{9}$$

$$\gamma_{gD}\mathcal{L}_{|g\rangle\langle e_D|}[\rho] \text{ (Decay of the dark excitonic level)}, \tag{10}$$

$$\gamma_{BD}\mathcal{L}_{|e_B\rangle\langle e_D|}[\rho] \text{ (Population transfer from the dark level into the bright level)}, \tag{11}$$

$$\gamma_{DB}\mathcal{L}_{|e_D\rangle\langle e_B|}[\rho] \text{ (Decay of the bright level into the dark level)}, \tag{12}$$

$$\gamma_{DD}\mathcal{L}_{|e_D\rangle\langle e_D|}[\rho] \text{ (Pure dephasing of the dark level)}, \tag{13}$$

$$\gamma_{BB}\mathcal{L}_{|e_B\rangle\langle e_B|}[\rho] \text{ (Pure dephasing of the bright level)}. \tag{14}$$

Furthermore, we assume that the process of population transfer from the dark state to the bright state (and vice versa) is thermally activated and hence we get:

$$\gamma_{DB} = [1 + N_{th}(\hbar\omega_B - \hbar\omega_D; T)]\gamma_{DB}^0, \tag{15}$$

$$\gamma_{BD} = N_{th}(\hbar\omega_B - \hbar\omega_D; T)\gamma_{DB}^0, \tag{16}$$

where $N_{th}(E;T)$ is the Bose-Einstein Distribution at temperature $T$ (we assume $T = 300$ K) and energy $E$ and $\gamma_{DB}^0$ is the spontaneous decay rate of the bright state, $|e_B\rangle$, into the dark state, $|e_D\rangle$. This rate, $\gamma_{DB}^0$, would be likely due to phonon-mediated processes that can lead to inter-exciton relaxation faster than the excitonic spontaneous emission [42]. The exact value of this rate would depend on the microscopic details of the electron-phonon interaction in a specific quantum dot, for which a valid estimate is very challenging to obtain [43-45].

In the PL calculations, we assume that the bright state, $|e_B\rangle$, as well as the dark state, $|e_D\rangle$, are incoherently pumped via terms $\gamma_{Bg}\mathcal{L}_{|e_B\rangle\langle g|}[\rho]$ and $\gamma_{Dg}\mathcal{L}_{|e_D\rangle\langle g|}[\rho]$, respectively. This accounts for pumping of the QD's states via another higher-energy bright state that is directly excited by an incident monochromatic laser. In the calculation of absorption and scattering, linear response theory is applied.

**Calculation of spectra.** We calculate the absorption, $S_{abs}(\omega)$, scattering, $S_{sca}(\omega)$, and emission, $S_{em}(\omega)$, spectra using the following formulas, valid close to the plasmonic resonance:

$$S_{abs}(\omega) \propto \omega Re\left\{\int_0^\infty \langle a(t)a^\dagger(0)\rangle e^{i\omega t}dt\right\}, \quad (17)$$

$$S_{sca}(\omega) \propto \omega^4 \left|\int_0^\infty \langle a(t)a^\dagger(0)\rangle e^{i\omega t}dt\right|^2, \quad (18)$$

$$S_{em}(\omega) \propto \omega^4 Re\left\{\int_0^\infty \langle a^\dagger(0)a(t)\rangle e^{i\omega t}dt\right\}. \quad (19)$$

Here we assume that the system absorbs, scatters and emits light predominantly via the plasmonic cavity and neglect any direct absorption, scattering or emission of the quantum dot.

We evaluate the two-time correlation functions $\langle a(0)a^\dagger(t)\rangle$ and $\langle a(t)a^\dagger(0)\rangle$ using the quantum regression theorem (QRT) as described elsewhere [46].

The second-order photon correlation function $g^{(2)}(t)$ is evaluated in the framework of cavity-quantum electrodynamics from the QRT as:

$$g^{(2)}(t) = \frac{\langle a^\dagger(0)a^\dagger(t)a(t)a(0)\rangle}{\langle a^\dagger a\rangle^2}. \quad (20)$$

**Analytical model for the excitonic photoluminescence spectrum.** We calculate the effective dynamics of the dark and the bright excitons with the use of an analytical model that we outline next. The explicit details can be found in the Supplementary Note. For the dark exciton, we eliminate the plasmonic cavity interacting with the bright exciton using the adiabatic approximation and obtain effective decay rates due to the cavity-induced Purcell effect $\gamma_{\text{Pur}}^{\text{D}}$. In this derivation, we use the separation of time scales present in the system due to the condition $g_{\text{B}} \gg g_{\text{D}}$ and show that the interaction of the dark exciton with the cavity is strongly influenced by the presence of the bright exciton, which significantly modifies the cavity response. The coupling of the bright exciton with the cavity, on the other hand, can be affected by the presence of the dark exciton via the incoherent pumping from the dark exciton, $\gamma_{\text{BD}}$, and in a smaller degree by the decay from the bright exciton into the dark one, $\gamma_{\text{DB}}$. As we show in Fig. 5, these incoherent processes barely influence the shape of the bright-exciton photoluminescence spectrum, but they can significantly contribute to its emission intensity. To derive the effective decay rate of the dark exciton and the steady-state populations of the dark exciton, $\langle\sigma_{\text{DD}}\rangle$, due to the cavity, we use the master equation to obtain a system of equations for the mean values of operators $\langle\sigma_{\text{DD}}\rangle$, $\langle a\sigma_{\text{gD}}^{\dagger}\rangle$, and $\langle\sigma_{\text{gD}}^{\dagger}\sigma_{\text{gB}}\rangle$, and apply the adiabatic approximation to obtain the steady states (see Supplementary Note).

We also decompose the emission spectra into the bright- and dark-state contributions. To that end we approximate the time evolution of the plasmon annihilation operator $a$ in the adiabatic approximation and express the emission spectrum in terms of the excitonic operators. In particular, we assume that we can split the total photoluminescence spectrum, $S_{\text{em}}(\omega) = S_{\text{em(B)}}(\omega) + S_{\text{em(D)}}(\omega)$, into the contributions that emerge due to the bright exciton, $S_{\text{em(B)}}(\omega)$, and due to the dark exciton, $S_{\text{em(D)}}(\omega)$, as displayed in Fig. 5(b). The details of the PL calculation for each of the contributions can be found in Supp. Mat.

**Data availability**

The datasets generated during and/or analysed during the current study are available from the corresponding author upon reasonable request.

**Code availability**

Codes used to generate simulated data are available from the corresponding author upon request.

**Acknowledgements**

SNG thanks the Government of Israel for a Planning and Budgeting Committee Fellowship. GH is the incumbent of the Hilda Pomeraniec Memorial Professorial Chair. RE, TN, and JA acknowledge funding from project FIS2016-80174-P of the Spanish Ministry of Science, Innovation and Universities MICINN, as well as funding from grant IT1164-19 for consolidated groups of the Basque University, through the Department of Universities of the Basque Government. This project received partial support from the European Union's Horizon 2020 research and innovation programme under grant agreement No 861950, project POSEIDON, and grant agreement no. 810626, project SINNCE.

**Author contributions**

S.N.G, O.B., T.N., R.E., J.A. and G.H designed the research. S.N.G., O.B. and G.H. designed and performed the experiments. T.N., R.E., and J.A. performed the theoretical analysis. L.C. helped in discussion and interpretation of the results. All authors participated in the analysis and discussion of the results, and in the writing of the manuscript.

**Competing interests**

The authors declare no competing interest.

**Figures**

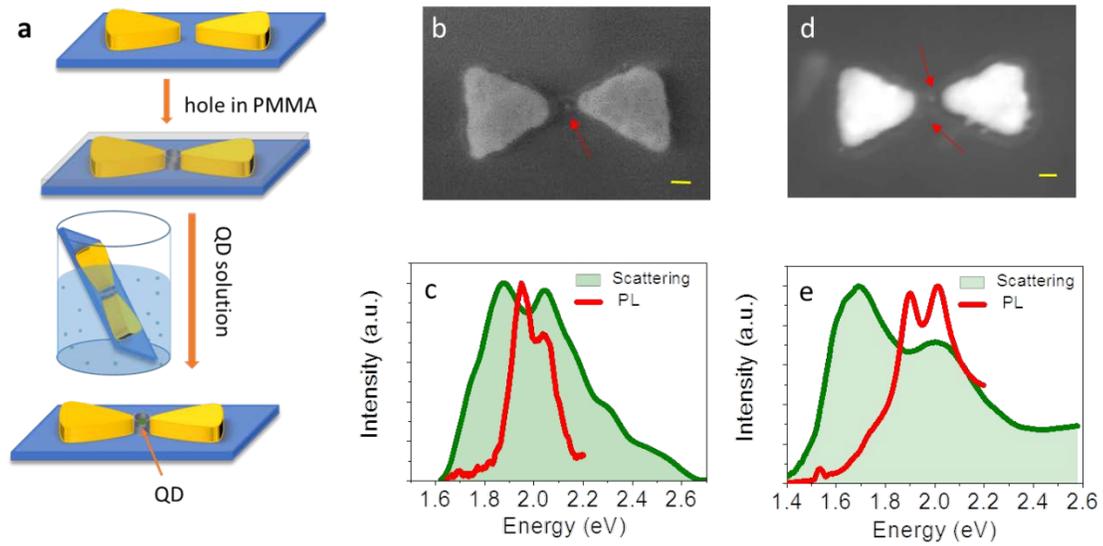

**Fig.1: Spectroscopy of plasmonic cavities with QDs.** (a) Schematic of the preparative process for trapping QDs within plasmonic bowties. (b,d) STEM images of a device with one QD (b) and two QDs (d). The scale bars represent 20 nm. The red arrows point to the QDs in the bowtie gaps. (c,e) Dark-field scattering spectra (green) and PL spectra (red) of the devices in b,d, respectively.

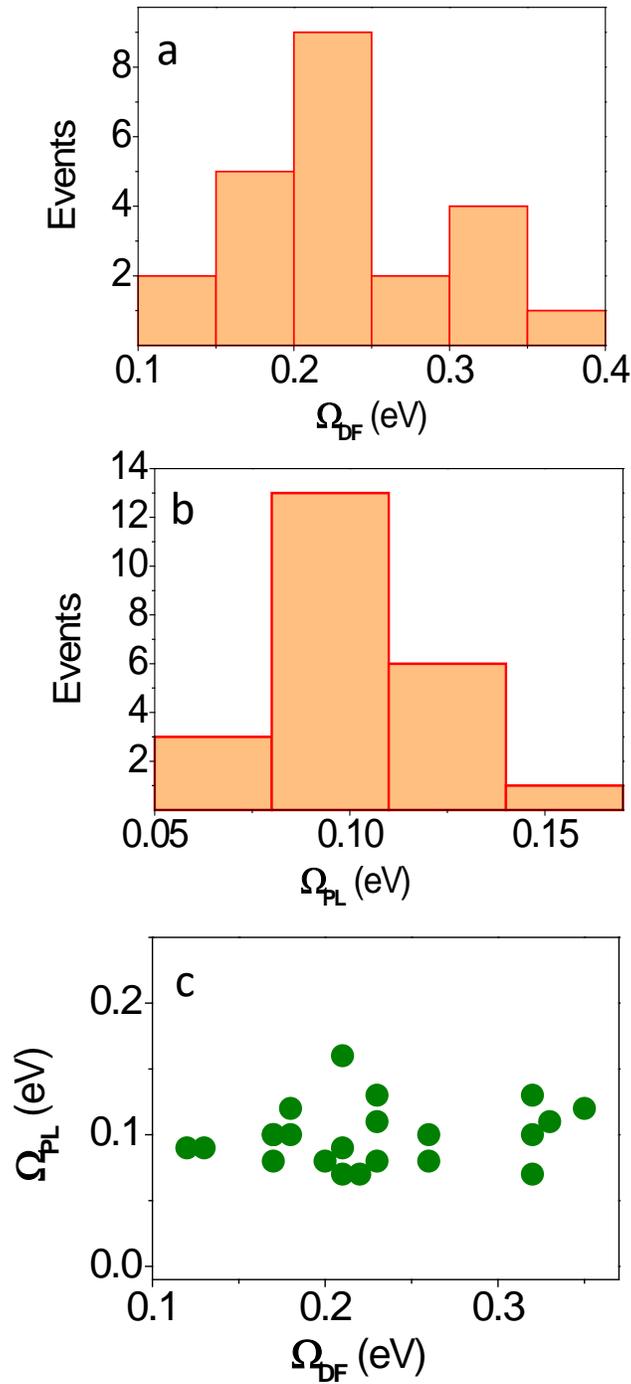

**Fig. 2: Peak splittings in scattering and PL.** (a, b) Histograms of peak splitting values obtained from dark-field scattering spectra, $\Omega_{DF}$ (a) and from PL spectra, $\Omega_{PL}$ (b). (c) Correlation between splitting values in PL and in scattering.

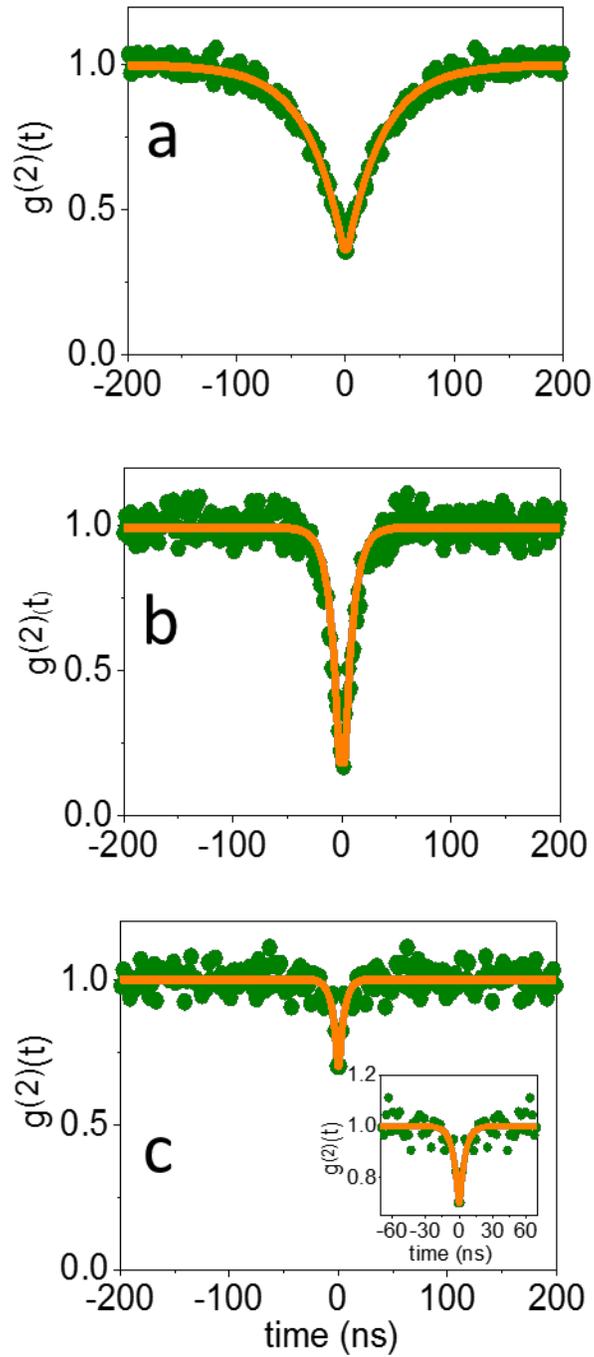

**Fig. 3: Second-order photon correlation function ($g^{(2)}(t)$).** (a) A bare QD on glass substrate. (b-c) QDs coupled to PCs. The dip at zero delay is a manifestation of a single QD in (b) and three QDs in, (c). Green- experimental results, orange- fits to equation 1.

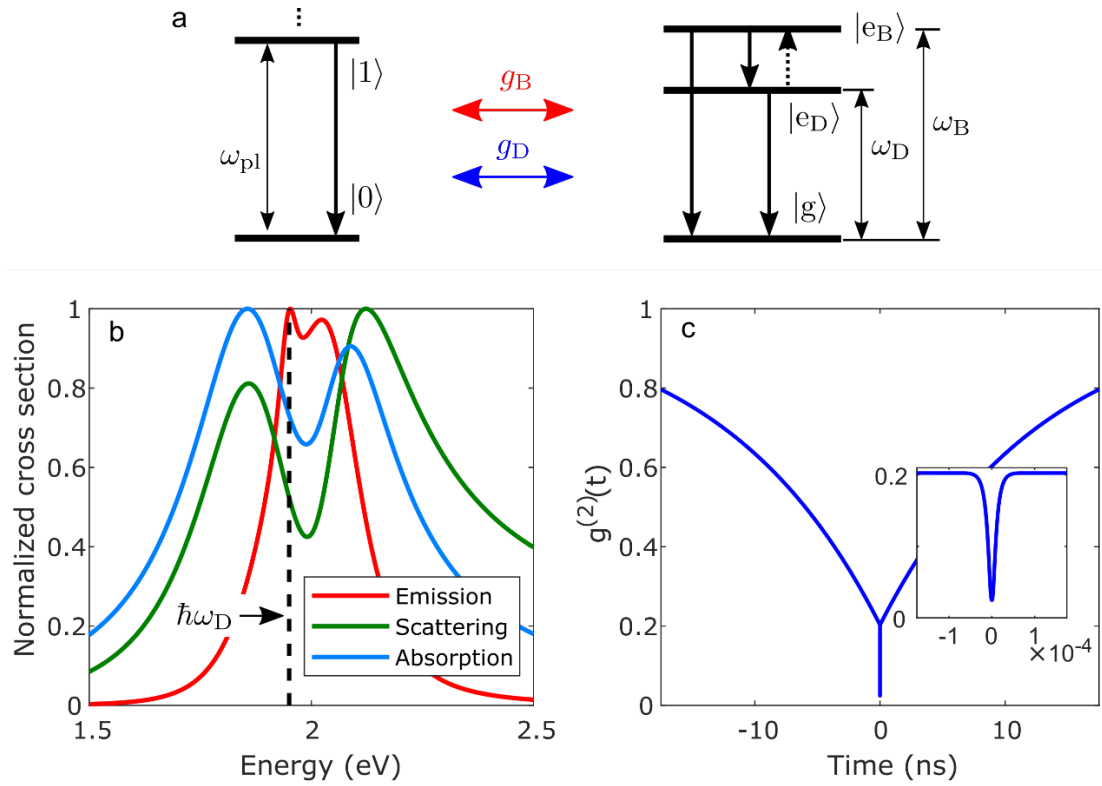

**Fig. 4: Quantum simulations of the plasmon-QD coupling dynamics.** (a) Schematic level diagram describing the theoretical model. The plasmonic cavity is depicted on the left with an excited state of energy $\omega_{pl}$. The QD (right) is described as a three-level electronic system containing a ground state |g⟩, a bright excitonic level, |e_B⟩, and a dark excitonic level, |e_D⟩. The bright (dark) excitonic transition occurs at energy $\omega_B(\omega_D)$. For the PL spectrum. we assume that both the bright and the dark excitons are pumped incoherently. Plasmon-exciton coupling is described within the Jaynes-Cummings model with rates $g_B$ and $g_D$ (for coupling of the bright and dark exciton, respectively). The arrows mark incoherent transfer rates considered in the model. (b) Emission (red), scattering (green), and absorption (blue) spectra calculated theoretically for parameters shown in Table 1. The dashed line marks the energy of the dark exciton, $\hbar\omega_D$. (c) A simulated $g^{(2)}(t)$ features a double-exponential decay. Inset shows a zoom of the fast (fs) decay of the system excitations that is not resolved on the ns time scale.

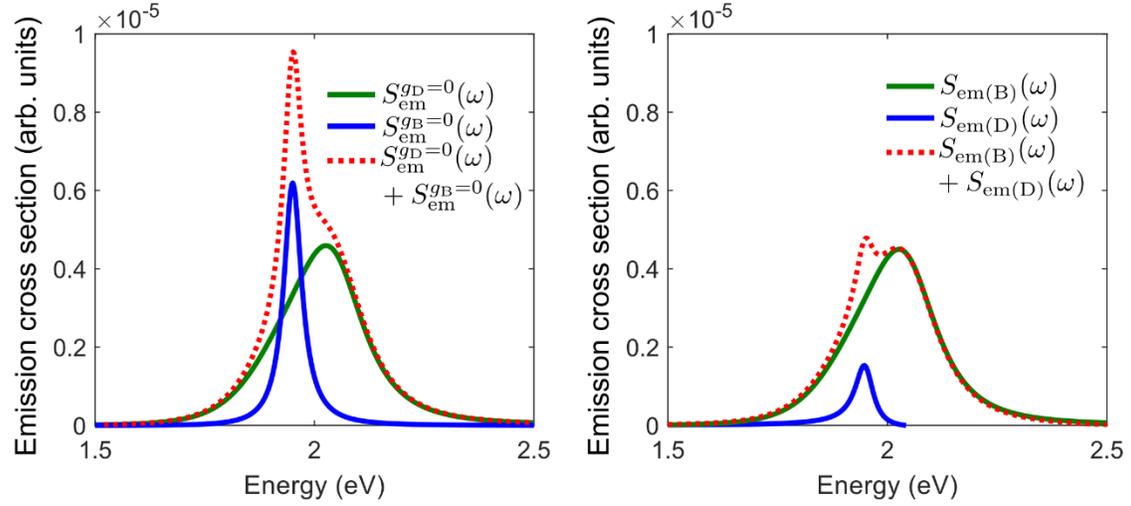

**Fig. 5: Quantum simulations of the plasmon-QD emission spectrum.** (a) Numerically calculated emission spectra of the system for $g_\text{D} = 0$, $S_\text{em}^{g_\text{D}=0}(\omega)$ (green line), for $g_\text{B} = 0$, $S_\text{em}^{g_\text{B}=0}(\omega)$ (blue line), and sum of the two, $S_\text{em}^{g_\text{D}=0}(\omega) + S_\text{em}^{g_\text{B}=0}(\omega)$ (red dotted line). (b) Analytically calculated contributions to the emission spectrum due to the bright exciton, $S_\text{em(B)}(\omega)$ (green line), dark exciton, $S_\text{em(D)}(\omega)$ (blue line), and the total analytical spectrum containing both contributions, $S_\text{em}(\omega) = S_\text{em(B)}(\omega) + S_\text{em(D)}(\omega)$ (red dotted line).

**Table 1. Set of parameters used to reproduce spectra and $g^{(2)}(t)$ functions, as shown in Fig. 4.** The choice of parameters has been guided by the experimental results.

| | | |
|---|---|---|
| $\hbar\kappa$ | 400 meV | Intrinsic plasmon decay rate |
| $\hbar\gamma_{gB}$ | 0.1 µeV | Intrinsic decay rate of the bright exciton |
| $\hbar\gamma_{gD}$ | 0.005 µeV | Intrinsic decay rate of the dark exciton |
| $\hbar\gamma_{DB}^0$ | 0.2 µeV | Rate of energy transfer between the dark and the bright exciton |
| $\hbar\gamma_{BB}$ | 130 meV | Pure dephasing of the bright exciton |
| $\hbar\gamma_{DD}$ | 50 meV | Pure dephasing of the dark exciton (Broadening of the dark-exciton line) |
| $\hbar\gamma_{Bg}$ | 1 neV | Incoherent pumping of the bright exciton |
| $\hbar\gamma_{Dg}$ | 5 neV | Incoherent pumping of the dark exciton |
| $\hbar g_B$ | 100 meV | Coupling between plasmon and the bright exciton |
| $\hbar g_D$ | 35 µeV | Coupling between plasmon and the dark exciton |
| $\hbar\omega_{pl}$ | 1.93 eV | Plasmon energy |
| $\hbar\omega_D$ | 1.95 eV | Emission energy of the dark exciton |
| $\hbar\omega_B$ | 2.00 eV | Emission energy of the bright exciton |

**Supplementary Information**

Supplementary Figures 1-11

Supplementary Table 1

Supplementary Note

# Supplementary Information

# Complex plasmon-exciton dynamics revealed through quantum dot light emission in a nanocavity

[1]Department of Chemical and Biological Physics, Weizmann Institute of Science, POB 26, Rehovot 7610001, Israel,

[2]Department of Chemical Research Support, Weizmann Institute of Science, POB 26, Rehovot 7610001, Israel,

[3]Materials Physics Center CSIC-UPV/EHU, Paseo Manuel de Lardizabal 5, 20018 Donostia-San Sebastián, Spain

[4]Donostia International Physics Center DIPC, Paseo Manuel de Lardizabal 4, 20018 Donostia-San Sebastián, Spain

[5]Schulich Faculty of Chemistry, Technion-Israel Institute of Technology, Haifa, Israel.

*Correspondence: gilad.haran@weizmann.ac.il, aizpurua@ehu.eus

[&]These authors contributed equally to the work.

# Supplementary Figures

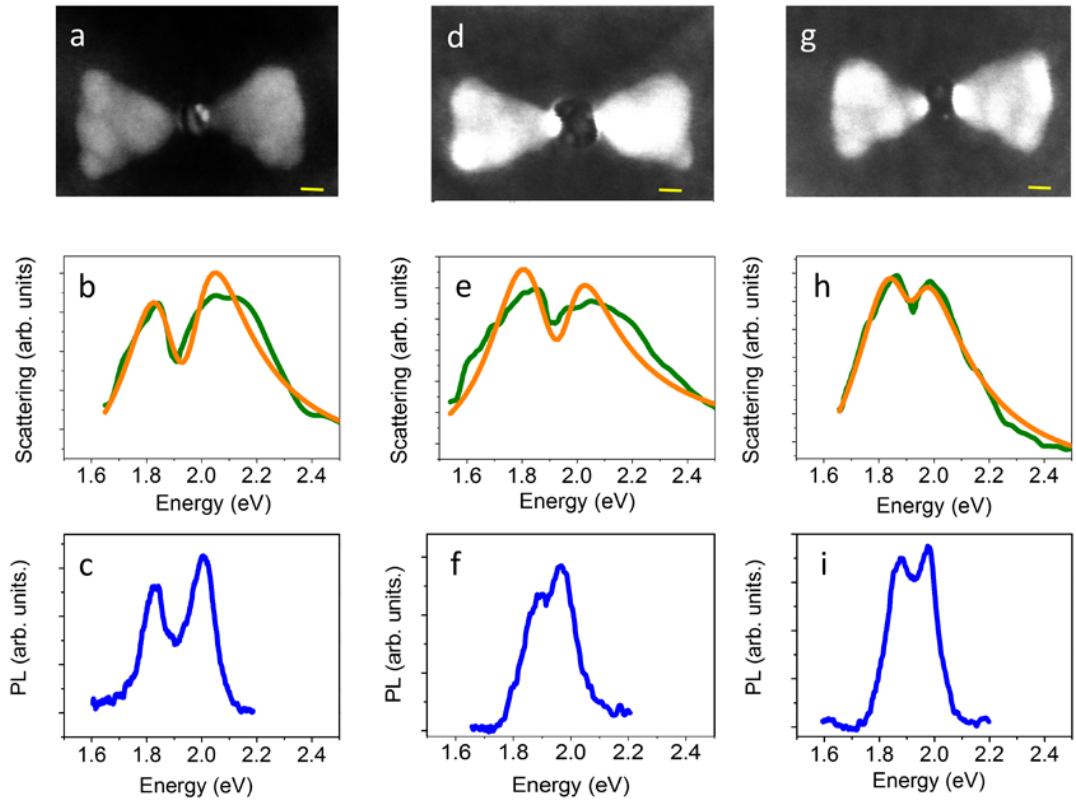

**Supplementary Figure 1: Additional examples of the spectroscopy of devices loaded with QDs.** STEM images (panels a,d,g), dark-field scattering spectra (panels b,e,h) and PL spectra (panels c,f,i) of three bowties containing QDs. The orange lines in panels b,e & h are fits to the coupled-oscillator model described in the legend of Supplementary Figure 2. The obtained coupling strengths are 81.3±0.6, 77.2±1.0 and 54.4±0.5 meV, respectively. Scale bars in panels a,d,g represent 20 nm.

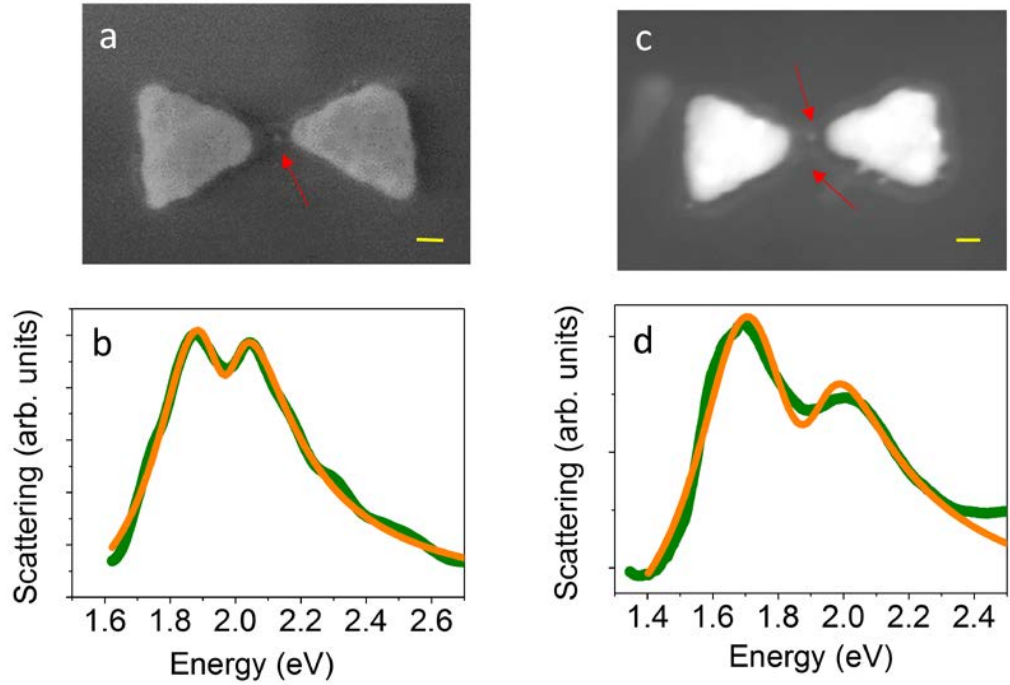

**Supplementary Figure 2: Coupled-oscillator model fits of scattering spectra:** STEM images (panels a&c) and dark-field scattering spectra (panels b&d) of two bowties containing QDs shown in Figure 1. The orange lines in panel b & d are fits to the coupled-oscillator model [27,28]:

$$S(\omega) \propto \omega^4 \left| \frac{(\omega_e^2 - \omega^2 - i\gamma_e \omega)}{(\omega^2 - \omega_p^2 + i\gamma_p \omega)(\omega^2 - \omega_e^2 + i\gamma_e \omega) - 4\omega^2 g^2} \right|^2,$$

where $\omega_e$ and $\gamma_e$ are the emitter resonance frequency and decay rate respectively, $\omega_p$ and $\gamma_p$ are the plasmon frequency and plasmon decay rate, respectively and g is the coupling rate. The obtained values of g are 52.6±0.3 and 103.5±1.1 meV for panels b and d, respectively. In these fits, as well as the fits in Supplementary Figure 1, we fixed the values of $\gamma_e$ (132 meV) and $\gamma_p$ (395 meV), based on the measured line widths of the individual QD PL and scattering spectra of the empty bowtie. A baseline parameter was used in the fitting in order to take care of a constant background signal.

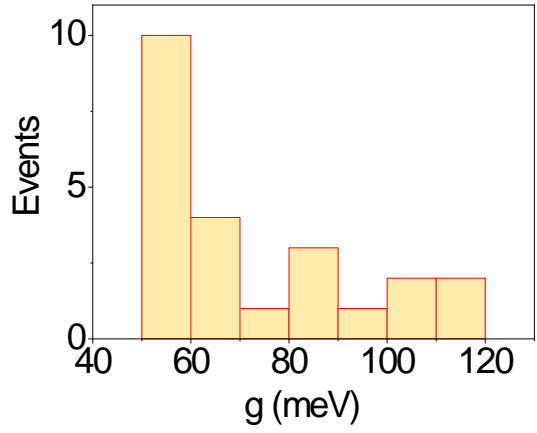

**Supplementary Figure 3: Coupling rate values.** Histogram shows the distribution of the values of coupling rates, g, obtained from fits of the scattering spectra using the coupled-oscillator model.

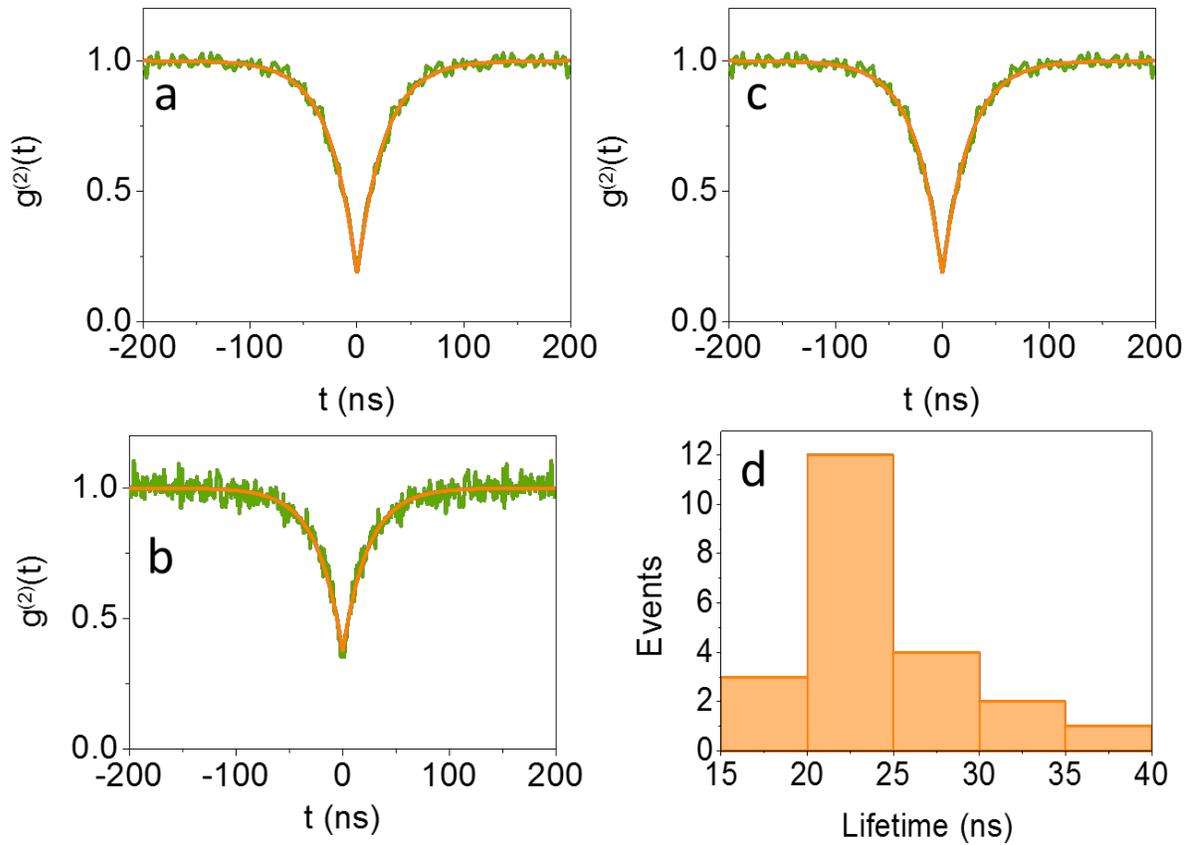

**Supplementary Figure 4: Second-order photon correlation function of QDs on glass.** (a-c) Three additional examples of measured correlation function of individual QDs on a glass substrate. (d) Distribution of the lifetimes of the excitons of 22 QDs on glass, obtained from analysis of $g^{(2)}$(t) functions.

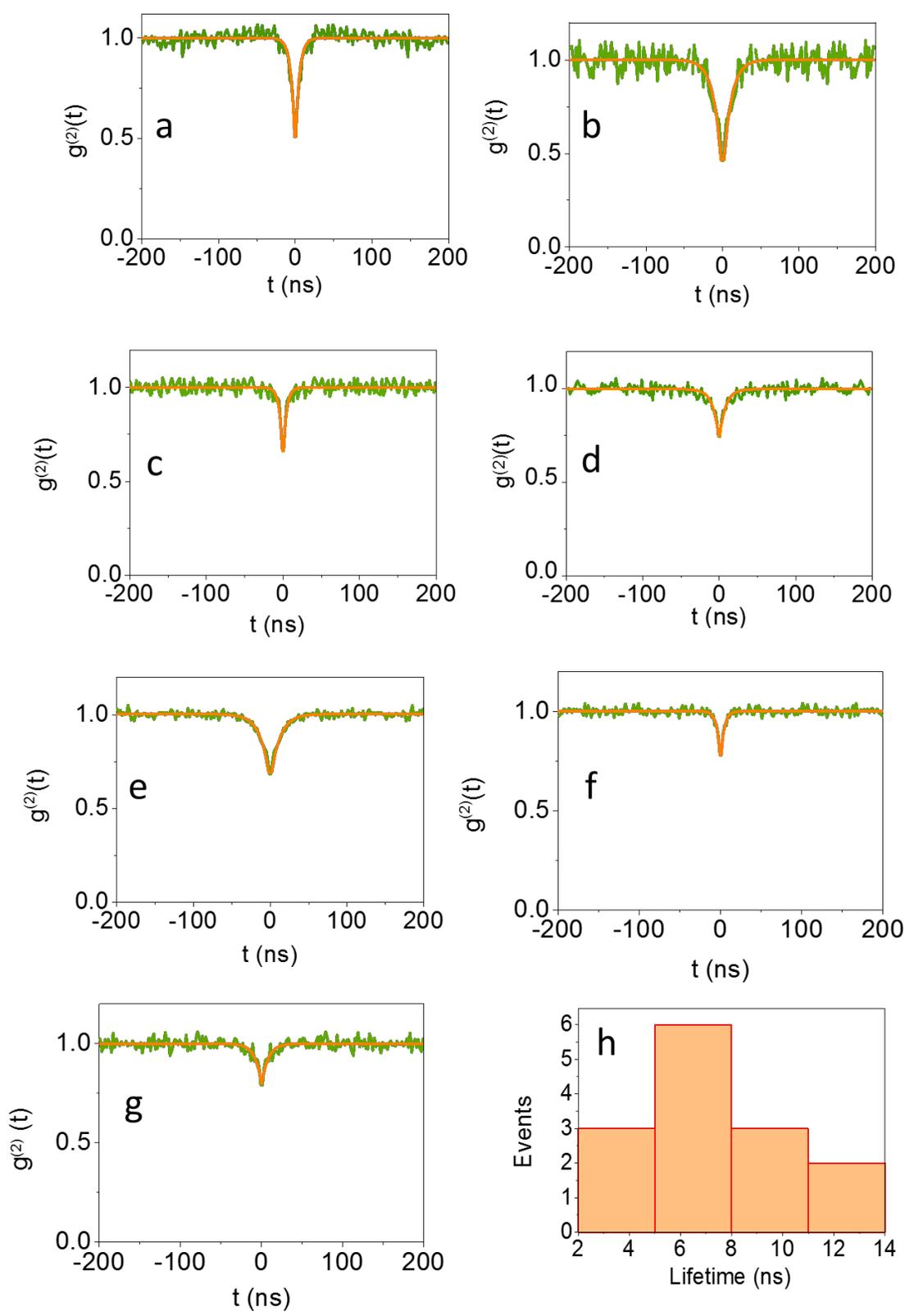

**Supplementary Figure 5: Second-order photon correlation function of QDs coupled to a PC.** (a-g) Additional examples of correlation functions measured from strongly coupled QDs. (h) Distribution of the lifetimes of 14 coupled plasmonic cavity-QD systems, obtained from analysis of $g^{(2)}$(t) functions.

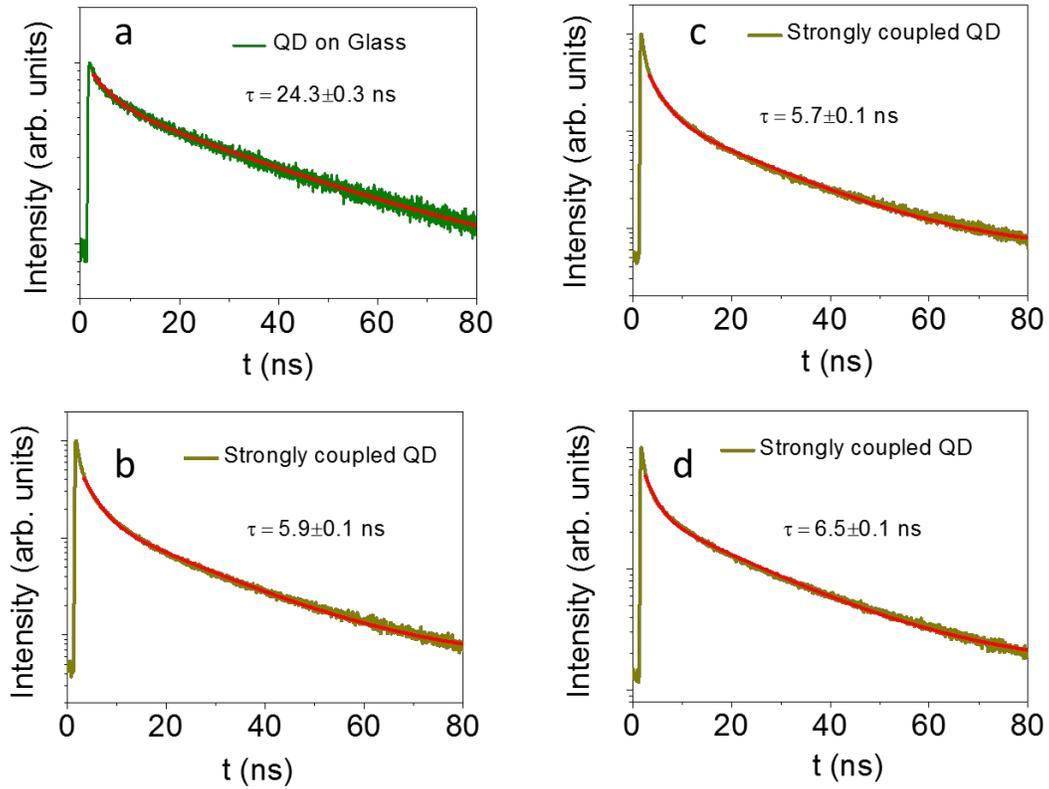

**Supplementary Figure 6: Time-resolved PL measurements of QDs.** (a) Measurement of QDs on a glass substrate. (b-d) Measurements of QDs coupled to PCs. The curves show the total decay of fluorescence, due to both radiative and non-radiative effects. The solid red lines are bi-exponential fits to the experimental data. The average lifetime (i.e. the weighted average of the two lifetimes obtained from the fits) is given in each panel.

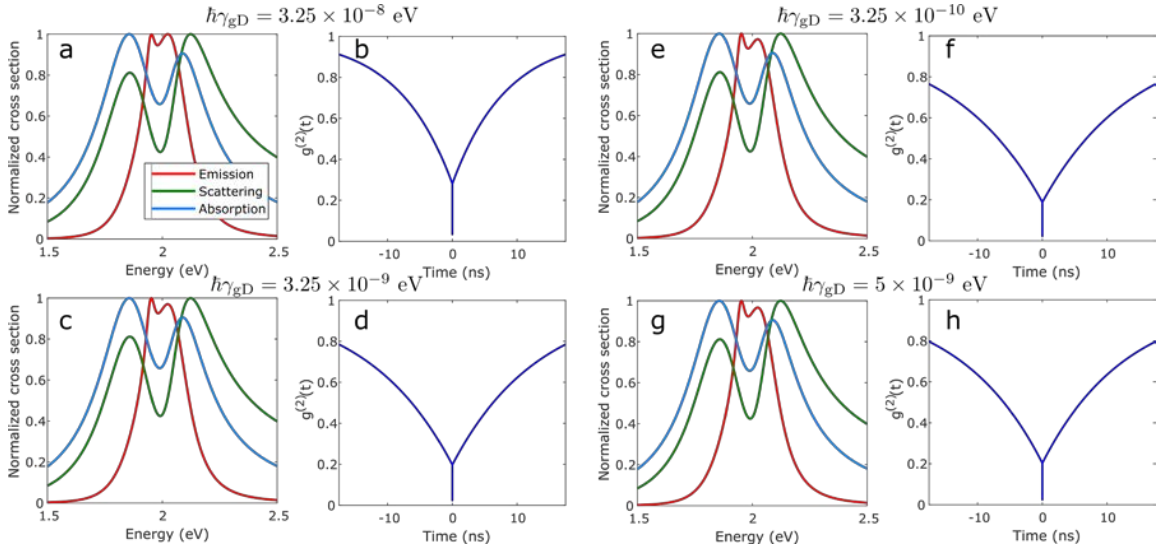

**Supplementary Figure 7: Dependence of spectra and correlation functions on the intrinsic decay rate of the dark exciton $\gamma_{gD}$.** A set of calculations of the emission, scattering and absorption of the antenna-emitter hybrid together with the correlation function of the emission, $g^{(2)}$, for situations in which the intrinsic decay rate of the dark exciton, $\gamma_{gD}$, is modified. The value of the dark exciton decay rate used is $\hbar\gamma_{gD}$= 3.25x10$^{-8}$ eV (a,b); $\hbar\gamma_{gD}$= 3.25x10$^{-9}$ eV (c,d); $\hbar\gamma_{gD}$= 3.25x10$^{-10}$ eV (e,f); and $\hbar\gamma_{gD}$= 5x10$^{-9}$ eV (g,h). The main features of the spectra and emission statistics are very robust against this parameter. This is due to the fact that the dynamics are governed by the rate of energy transfer between the dark and the bright exciton rather that the intrinsic dark exciton lifetime itself.

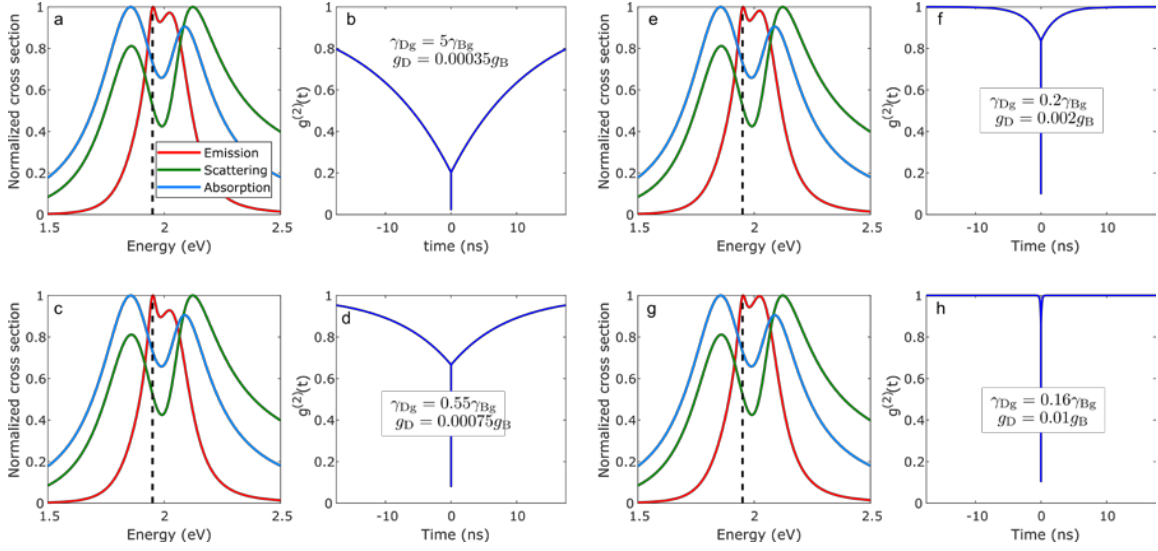

**Supplementary Figure 8: Influence of the model parameters on simulated spectra and correlation functions.** Model parameters are modified here in order to test how they affect predictions for experimental observables. (a,c,e,g) Emission (red line), absorption (blue line), and scattering (green line) spectra and (b,d,f,h) $g^{(2)}(t)$ functions calculated from the theoretical model considering different ratios of incoherent pumping ($\gamma_{Bg}/\gamma_{Dg}$, with $\hbar\gamma_{Bg} = 1$ neV in all cases) and different values of the Jaynes-Cummings coupling constant of the dark exciton with plasmons, $g_D$. The values are given in the inset and apply to each pair [(a,b), (c,d), (e,f), and (g,h)] separately. All other parameters are listed in Table 1. The spectral response is almost identical in all cases, but the form of $g^{(2)}(t)$ very sensitively depends on both $\gamma_{Bg}/\gamma_{Dg}$ and g$_D$. The ratio $\gamma_{Bg}/\gamma_{Dg}$ controls the relative contribution of the fast component of the decay with respect to the slow one. On the other hand, large values of $g_D$ give rise to a Purcell effect that shortens the lifetime of the dark exciton, thus shortening the lifetime associated with the slow decay. By fitting an exponential function of the form $f(t) = 1 - Ae^{-\frac{t}{T}}$ to the slow-decaying tails of the correlation function $g^{(2)}(t)$ we have obtained the following lifetimes of the dark exciton: (b) T=12.8 ns, (d) T=8.9 ns, (f) T=2.3 ns, (h) T=0.1 ns.

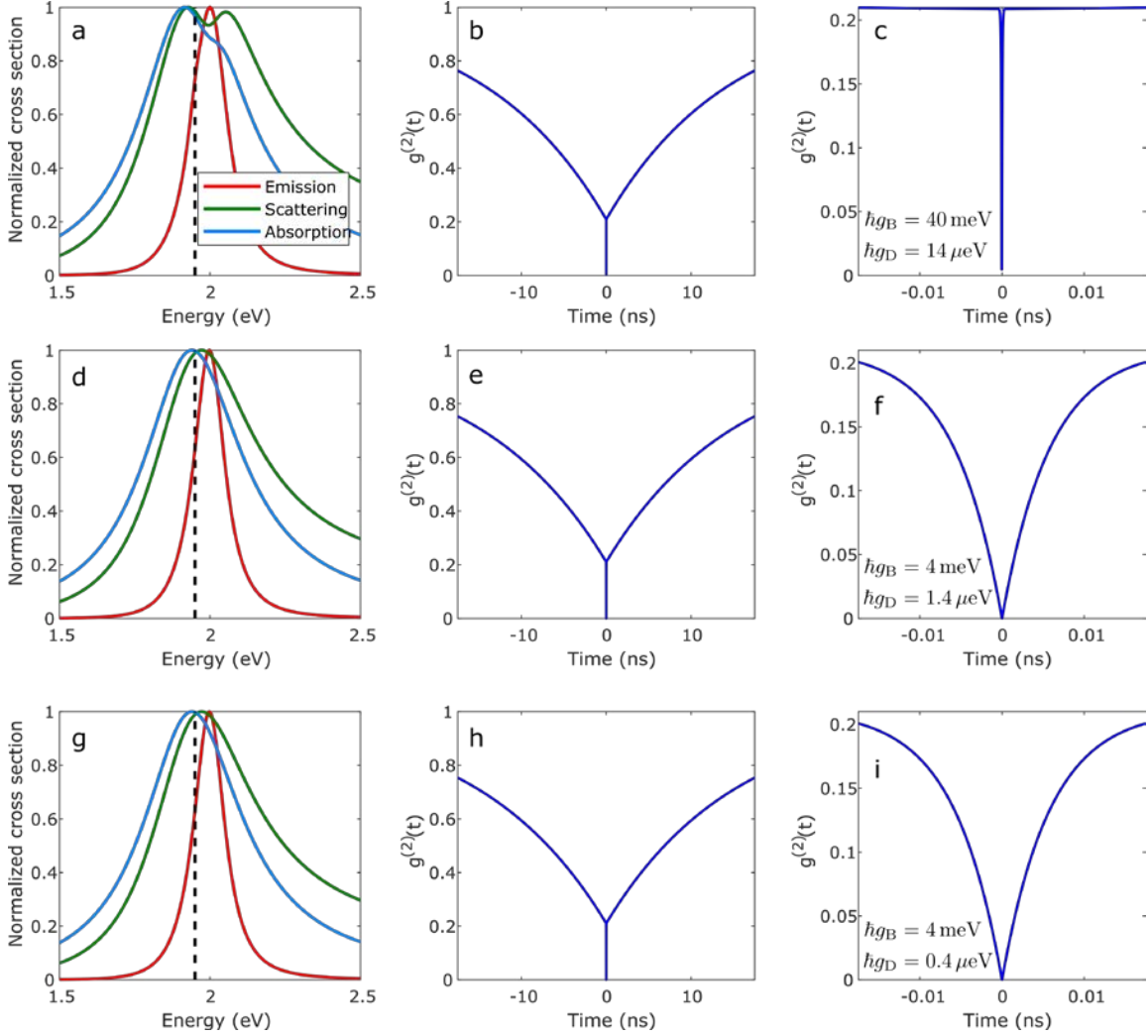

**Supplementary Figure 9: Influence of the plasmon-exciton couplings on spectra and on the time resolution of the fast component in photon correlation functions.** Optical properties of emission, scattering, and absorption (a,d,g) of the antenna-emitter hybrid for different parameters of the bright and dark exciton coupling to the antenna, $g_B$ and $g_D$, respectively. The second-order correlation function of the emission g$^{(2)}$ for the same coupling strengths are displayed in (b,e,h). A zoom-in of the correlation function in a shorter time scale is provided in (c,f,i). The coupling parameters considered are $\hbar g_B$= 40 meV; $\hbar g_D$= 14 μeV in (a,b,c); $\hbar g_B$= 4 meV; $\hbar g_D$= 1.4 μeV in (d,e,f); and $\hbar g_B$= 4 meV; $\hbar g_D$= 0.4 μeV in (g,h,i). One can observe that when the coupling of the bright state with the plasmon is decreased to 4 meV (f,i), the fast component of the biexponential decay becomes slower, giving rise to a broader dip in the correlation function, of the order of approximately 10 picoseconds, which might bring the observation of this fast-decaying component to the edge of experimental

observation. However, a coupling of 4 meV is in the weak coupling regime, and therefore no fingerprint of strong coupling would be observable any more under these conditions.

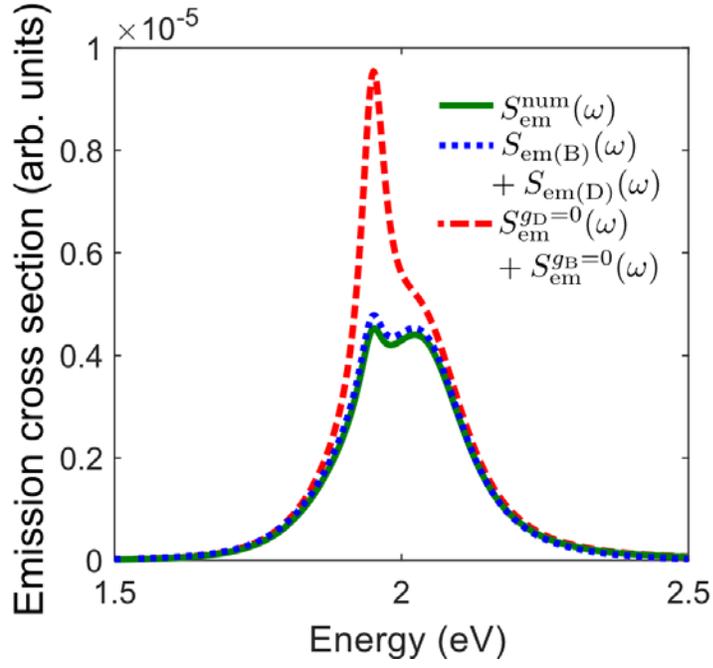

**Supplementary Figure 10: Comparison of the PL spectrum obtained from the numerical calculation and from the analytical decomposition**. The result of a full numerical calculation of the emission spectra (green line) is compared with the analytical result as in Fig. 5(b) of the main text (blue dotted line), and with the sum of numerically calculated spectra $S_{\text{em}}^{g_D=0}(\omega) + S_{\text{em}}^{g_B=0}(\omega)$ (red dashed line), as in Fig. 5(a) of the main text. All spectra are calculated for the set of parameters specified in Table 1.

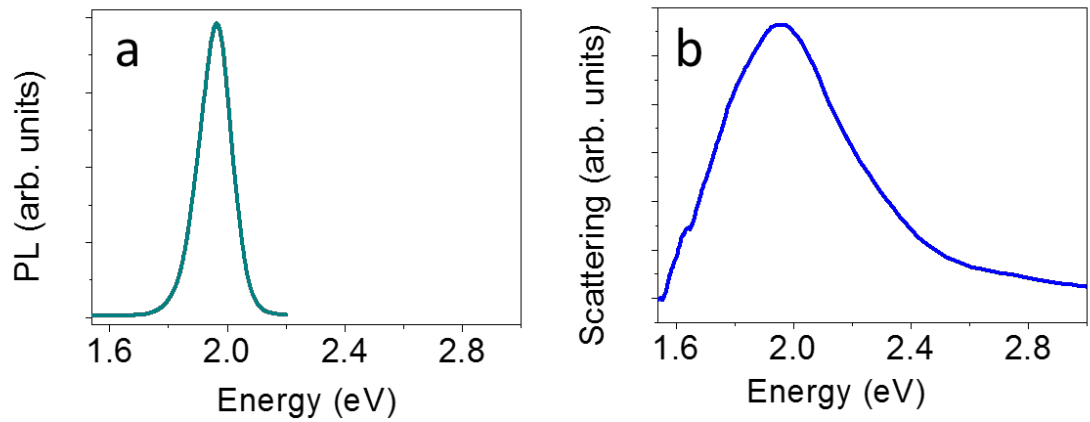

**Supplementary Figure 11: Spectroscopy of a bare QD and an empty plasmonic cavity.** (a) PL spectrum of a bare QD on a glass substrate. (b) Dark-field scattering spectrum of an empty plasmonic bowtie.

# Supplementary Tables

Supplementary Table 1: Decay times and amplitudes extracted from biexponential fits to time-resolved PL curves.

|              | $a_1$ | $\tau_1$ (ns) | $a_2$ | $\tau_2$ (ns) | $\langle\tau\rangle$ (ns) |
|--------------|-------|---------------|-------|---------------|----------------------------|
| **QD on glass**  | 0.43  | 4.33          | 0.57  | 39.16         | 24.18                      |
| **Coupled QD 1** | 0.82  | 2.91          | 0.18  | 19.09         | 5.82                       |
| **Coupled QD 2** | 0.82  | 2.89          | 0.18  | 18.51         | 5.70                       |
| **Coupled QD 3** | 0.75  | 2.30          | 0.25  | 19.55         | 6.61                       |

$a_1$ and $a_2$ are amplitudes and $\tau_1$ and $\tau_2$ are decay times, while $\langle\tau\rangle$ is the average decay time.

# Supplementary Note: Analytical treatment of the plasmon-exciton dynamics and photoluminescence.

To obtain an analytical insight into the exciton dynamics and the origin of the features appearing in the excitonic light-emission spectra, we use the separation of time scales that naturally emerges in the system under study. The bright exciton is coupled to the single mode of the plasmonic structure with a coupling strength $g_\text{B}$ comparable to the intrinsic losses $\kappa$ of the plasmon, and therefore the dynamics of the plasmon and the bright exciton must be treated on the same footing. The dark exciton, however, is coupled to the plasmon with a coupling strength $g_\text{D} \ll g_\text{B}$, which allows us to treat this coupling perturbatively. This separation of time scales allows us to obtain an effective cavity-induced decay rate of the dark exciton $\gamma_\text{Pur}^\text{D}$ (Purcell effect) and the resulting dark-exciton steady-state population. Additional incoherent coupling between the dark exciton and the bright exciton with rates $\gamma_\text{BD}$ (dark-to-bright-exciton incoherent coupling rate) and $\gamma_\text{DB}$ (bright-to-dark-exciton incoherent coupling rate) are also included in the model. As we demonstrate below, their role is mainly to influence the relative distribution of excitonic population between the bright and the dark exciton and thus rescale the amplitudes of the features in the emission spectra arising from the bright- and the dark-exciton emission, respectively.

### *Effective decay rate of the dark exciton*

In this section we provide details on the calculation of the effective dynamics (decay) of the dark exciton.

We calculate the effective decay rate of the dark exciton interacting with the cavity using the master equation. We first obtain approximate expressions for the following operator mean values:

$$\frac{d}{dt}\langle \sigma_\text{DD}\rangle = -(\gamma_\text{gD} + \gamma_\text{BD})\langle \sigma_\text{DD}\rangle - ig_\text{D}\left\langle a\sigma_\text{gD}^\dagger\right\rangle + ig_\text{D}\langle a^\dagger \sigma_\text{gD}\rangle + \gamma_\text{Dg}, \qquad (1)$$

$$0 \approx \frac{d}{dt}\left\langle a\sigma_\text{gD}^\dagger\right\rangle = \left[-i(\omega_\text{pl} - \omega_\text{D}) - \left(\frac{\Gamma_\text{D}}{2} + \frac{\kappa}{2}\right)\right]\left\langle a\sigma_\text{gD}^\dagger\right\rangle - ig_\text{D}\langle \sigma_\text{DD}\rangle - ig_\text{B}\left\langle \sigma_\text{gD}^\dagger \sigma_\text{gB}\right\rangle, \qquad (2)$$

$$0 \approx \frac{d}{dt}\left\langle \sigma_{gD}^{\dagger}\sigma_{gB} \right\rangle = \left[-i(\omega_B - \omega_D) - \left(\frac{\Gamma_D}{2} + \frac{\Gamma_B}{2}\right)\right]\left\langle \sigma_{gD}^{\dagger}\sigma_{gB} \right\rangle - ig_B\left\langle a\sigma_{gD}^{\dagger} \right\rangle, \quad (3)$$

with $\sigma_{gB} = |g\rangle\langle e_B|$, $\sigma_{gD} = |g\rangle\langle e_D|$ and $\sigma_{DD} = |e_D\rangle\langle e_D|$. On the right-hand side of Supplementary Eq. (2) and (3) we have neglected $g_D\langle a^{\dagger}\sigma_{gB}\rangle \approx 0$ and $g_D\langle a^{\dagger}a\rangle \approx 0$, and used the adiabatic approximation to set the time derivatives in Supplementary Eqs. (2) and (3) equal to zero. From Supplementary Eqs. (2-3) we obtain:

$$\left\langle a\sigma_{gD}^{\dagger} \right\rangle \approx \frac{ig_D\left[\frac{(\gamma_B + \Gamma_D)}{2} + i(\omega_B - \omega_D)\right]\langle \sigma_{DD} \rangle}{g_B^2 + \left[\frac{(\gamma_B + \Gamma_D)}{2} + i(\omega_B - \omega_D)\right]\left[\frac{\kappa}{2} + i(\omega_{pl} - \omega_D)\right]}. \quad (4)$$

Supplementary Eqs. (1-3) also yield an effective equation of motion for $\langle \sigma_{DD} \rangle$ upon insertion of the approximate steady-state solution for $\left\langle a\sigma_{gD}^{\dagger} \right\rangle$ and $\langle a^{\dagger}\sigma_{gD}\rangle = \left\langle a\sigma_{gD}^{\dagger} \right\rangle^{*}$:

$$\frac{d}{dt}\langle \sigma_{DD} \rangle \approx -(\gamma_{gD} + \gamma_{BD} + \gamma_{Pur}^{D})\langle \sigma_{DD} \rangle + \gamma_{Dg},$$

where

$$\gamma_{Pur}^{D} \approx \frac{g_D^2\left[g_B^2(\Gamma_B + \Gamma_D) + (\kappa + \Gamma_D)\left(\frac{(\Gamma_B + \Gamma_D)^2}{4} + (\omega_B - \omega_D)^2\right)\right]}{g_B^4 + 2g_B^2 D_1 + D_2}, \quad (5)$$

and

$$D_1 = \left[\frac{(\kappa + \Gamma_D)(\Gamma_B + \Gamma_D)}{4} - (\omega_{pl} - \omega_D)(\omega_B - \omega_D)\right],$$

$$D_2 = \left[(\omega_{pl} - \omega_B)^2 + \frac{(\kappa + \Gamma_D)^2}{4}\right]\left[(\omega_D - \omega_B)^2 + \frac{(\Gamma_B + \Gamma_D)^2}{4}\right].$$

With the effective decay rate in hand, we can obtain the steady-state populations of the dark exciton, $\langle \sigma_{DD} \rangle$:

$$\langle \sigma_{DD} \rangle \approx \frac{\gamma_{Dg}}{\Gamma_{n_D}}, \quad (6)$$

Where $\Gamma_{n_D} = \gamma_{gD} + \gamma_{BD} + \gamma_{Pur}^D$ and we have assumed that the population of the bright state $\langle \sigma_{BB} \rangle$ is small so that $\gamma_{gD} \gg \gamma_{DB}\langle \sigma_{BB} \rangle$ (with $\sigma_{BB} = |e_B\rangle\langle e_B|$).

### *Decomposition of the emission spectra into the bright- and dark-state contribution*

Next, we calculate the approximate photoluminescence spectrum of the light emitted from the plasmonic cavity and directly link the light emission to the underlying excitonic dynamics. To that end, we approximate the time evolution of the plasmon annihilation operator $a$ in the adiabatic approximation and express the emission spectrum in terms of the excitonic operators. In particular, we assume that we can split the total photoluminescence spectrum, $S_{em}(\omega) = S_{em(B)}(\omega) + S_{em(D)}(\omega)$, into the contributions that emerge due to the bright exciton, $S_{em(B)}(\omega)$, and dark exciton, $S_{em(D)}(\omega)$, respectively:

$$S_{em(B)}(\omega) = 2\omega^4 Re\left\{\int_0^\infty \langle a_B^\dagger(0) a_B(t) \rangle e^{i\omega t} dt \right\}, \quad (7)$$

$$S_{em(D)}(\omega) = 2\omega^4 Re\left\{\int_0^\infty \langle a_D^\dagger(0) a_D(t) \rangle e^{i\omega t} dt \right\}. \quad (8)$$

Here we have used the lower index B and D to explicitly mark the dynamics of the bright and the dark exciton, respectively. For brevity we omit this index in the following discussion of the individual spectral contributions, unless it is needed for clarity.

### *Photoluminescence spectrum due to the bright exciton, $S_{em(B)}(\omega)$*

To obtain $S_{em(B)}(\omega)$ we use the quantum regression theorem (QRT), assuming again that the dark exciton does not significantly influence the dynamics of the coupling between the plasmon and the bright exciton:

$$\frac{d}{dt}\langle a^\dagger(0)a(t)\rangle = \left(-i\omega_{pl} - \frac{\kappa}{2}\right)\langle a^\dagger(0)a(t)\rangle - ig_B \langle a^\dagger(0)\sigma_{gB}(t)\rangle,$$

$$\frac{d}{dt}\langle a^\dagger(0)\sigma_{gB}(t)\rangle = \left(-i\omega_B - \frac{\Gamma_B}{2}\right)\langle a^\dagger(0)\sigma_{gB}(t)\rangle - ig_B \langle a^\dagger(0)a(t)\rangle.$$

After inserting the result into Supplementary Eq. (7) we obtain:

$$S_{em(B)}(\omega) = 2\omega^4 Re \left\{ \frac{ig_B\langle a^\dagger \sigma_{gB}\rangle + i\left(\omega - \omega_B + \frac{i\Gamma_B}{2}\right)\langle a^\dagger a\rangle}{\left(\omega - \omega_B + \frac{i\Gamma_B}{2}\right)\left(\omega - \omega_{pl} + \frac{i\kappa}{2}\right) - g_B^2} \right\}, \quad (9)$$

with

$$\langle a^\dagger \sigma_{gB} \rangle = \frac{g_B \kappa [\omega_B - \omega_{pl} + i(\kappa + \Gamma_B)/2\ ](\gamma_{Bg} + \gamma_{BD}\langle \sigma_{DD} \rangle)}{g_B^2 (\Gamma_B + \kappa)(\gamma_B + \kappa) + \gamma_B \kappa \left[(\omega_B - \omega_{pl})^2 + \frac{\kappa + \Gamma_B}{4}\right]},$$

$$\langle a^\dagger a \rangle = \frac{g_B^2 (\Gamma_B + \kappa)(\gamma_{Bg} + \gamma_{BD}\langle \sigma_{DD} \rangle)}{g_B^2 (\Gamma_B + \kappa)(\gamma_B + \kappa) + \gamma_B \kappa \left[(\omega_B - \omega_{pl})^2 + \frac{\kappa + \Gamma_B}{4}\right]},$$

Where $\gamma_B = \gamma_{gB} + \gamma_{DB}$ (i.e. it does not contain pure dephasing processes but only decay processes). The expressions for $\langle a^\dagger a \rangle$, $\langle a^\dagger \sigma_{gB} \rangle$, $\langle \sigma_{BB} \rangle$ were obtained from the following system of steady-state equations derived from the master equation:

$$\kappa \langle a^\dagger a \rangle + i g_B \langle a^\dagger \sigma_{gB} \rangle - i g_B \langle a \sigma_{gB}^\dagger \rangle = 0, \tag{10}$$

$$\gamma_B \langle \sigma_{BB} \rangle + i g_B \langle a \sigma_{gB}^\dagger \rangle - i g_B \langle a^\dagger \sigma_{gB} \rangle = \gamma_{Bg} + \gamma_{BD}\langle \sigma_{DD} \rangle, \tag{11}$$

$$\left[i(\omega_{pl} - \omega_B) - \left(\frac{\Gamma_B}{2} + \frac{\kappa}{2}\right)\right]\langle a \sigma_{gB}^\dagger \rangle - i g_B (\langle a^\dagger a \rangle - \langle \sigma_{BB} \rangle) = 0, \tag{12}$$

$$\left[i(\omega_B - \omega_{pl}) - \left(\frac{\Gamma_B}{2} + \frac{\kappa}{2}\right)\right]\langle a^\dagger \sigma_{gB} \rangle + i g_B (\langle a^\dagger a \rangle - \langle \sigma_{BB} \rangle) = 0. \tag{13}$$

### *Emission spectrum due to the dark exciton $S_{em(D)}(\omega)$*

The contribution to the emission spectrum arising due to the dark state can be obtained using the eigenvector perturbation theory to approximate the two-time correlation function $\langle a^\dagger(0) a(t) \rangle$. We can obtain from the QRT the following system of differential equations for the two-time correlation functions:

$$\frac{d}{dt}\begin{bmatrix} \langle a^\dagger(0)a(t)\rangle \\ \langle a^\dagger(0)\sigma_{gB}(t)\rangle \\ \langle a^\dagger(0)\sigma_{gD}(t)\rangle \end{bmatrix} =$$

$$\left( \underbrace{\begin{bmatrix} -i\omega_{pl} - \frac{\kappa}{2} & -ig_B & 0 \\ -ig_B & -i\omega_B - \frac{\Gamma_B}{2} & 0 \\ 0 & 0 & -i\omega_D - \frac{\Gamma_D}{2} \end{bmatrix}}_{M_0} + \underbrace{\begin{bmatrix} 0 & 0 & -ig_D \\ 0 & 0 & 0 \\ -ig_D & 0 & 0 \end{bmatrix}}_{\delta M} \right) \begin{bmatrix} \langle a^\dagger(0)a(t)\rangle \\ \langle a^\dagger(0)\sigma_{gB}(t)\rangle \\ \langle a^\dagger(0)\sigma_{gD}(t)\rangle \end{bmatrix}, \quad (14)$$

The first matrix in the parenthesis on the right-hand side, denoted as $M_0$, is responsible for the dynamics of the unperturbed system, whereas the second matrix in the parenthesis, denoted as $\delta M$, represents the perturbative coupling of the dark exciton to the strongly interacting system composed by the bright exciton and the plasmon. The differential equation can be formally solved using the eigenvalue decomposition of matrix $M = M_0 + \delta M$. If $M$ has non-degenerate eigenvalues $\lambda_i$ and corresponding left (right) eigenvectors $y_i$ ($x_i$), the solution of the differential equation is:

$$A(t) = \sum_i c_i x_i e^{\lambda_i t},$$

with

$$A(t) = \begin{bmatrix} \langle a^\dagger(0)a(t)\rangle \\ \langle a^\dagger(0)\sigma_{gB}(t)\rangle \\ \langle a^\dagger(0)\sigma_{gD}(t)\rangle \end{bmatrix},$$

and the coefficients $c_i$ are determined from the initial condition:

$$A(0) = \begin{bmatrix} \langle a^\dagger a\rangle \\ \langle a^\dagger \sigma_{gB}\rangle \\ \langle a^\dagger \sigma_{gD}\rangle \end{bmatrix},$$

as

$$\begin{bmatrix} c_1 \\ c_2 \\ c_3 \end{bmatrix} = \begin{bmatrix} y_{1,1} & y_{1,2} & y_{1,3} \\ y_{2,1} & y_{2,2} & y_{2,3} \\ y_{3,1} & y_{3,2} & y_{3,3} \end{bmatrix} \begin{bmatrix} \langle a^\dagger a\rangle \\ \langle a^\dagger \sigma_{gB}\rangle \\ \langle a^\dagger \sigma_{gD}\rangle \end{bmatrix},$$

where $y_{i,j}$ is the $j$-th component of the $i$-th left eigenvector. In the perturbative approach we first obtain the exact eigenvectors $x_{0i}$ and $y_{0i}$ of $M_0$. We further assume that the eigenvectors $x_{01}, x_{02}$ ($y_{01}, y_{02}$) belong to the subspace describing the dynamics of the bright exciton interacting with the plasmon, and the vector $x_{03}$ ($y_{03}$) belongs to the dark-exciton subspace of $M_0$. The approximate eigenvectors of $M$ can be found as:

$$x_i \approx \sum_{ij} \epsilon_{ij} x_{0j},$$

where

$$\epsilon_{ij} = \frac{y_{0j}^T \delta M x_{0i}}{(\lambda_{0i} - \lambda_{0j}) y_{0j}^T x_{0j}}.$$

To obtain the approximate emission spectrum we further assume that the eigenvectors corresponding to the dynamics of the plasmon coupled with the bright exciton, $x_1, x_2$ ($y_1, y_2$), remain approximately unchanged (decoupled from the dark exciton):

$$x_1 \approx x_{01}, x_2 \approx x_{02},$$

but we apply perturbation theory to obtain the eigenvector $x_3$:

$$x_3 \approx \sum_i \epsilon_{3i} x_{0i}.$$

The solution then separates into two independent contributions that give rise to: (i) the emission from the bright exciton coupled with the plasmon, $\langle a_B^\dagger(0) a_B(t) \rangle = \sum_{i=1,2} c_i x_{i,1} e^{\lambda_i t}$ (corresponding to the spectrum $S_{em(B)}(\omega)$ shown above for $\lambda_i \approx \lambda_{0i}$, with the unperturbed eigenvalues $\lambda_{0i}$ and where $x_{i,j}$ is the $j$-th component of the $i$-th right eigenvector), and (ii) the emission due to the dark exciton, $\langle a_D^\dagger(0) a_D(t) \rangle = c_3 x_{3,1} e^{\lambda_3 t}$. After performing the algebraic manipulations and inserting the result into Supplementary Eq. (8) we obtain:

$$S_{em(D)}(\omega) \approx 2\omega^4 \text{Re}\left\{\frac{B}{i(\omega_D - \omega) + \frac{\Gamma_D}{2}}\right\}, \qquad (15)$$

with

$$B = \frac{ig_D\left[\frac{(\gamma_B - \Gamma_D)}{2} + i(\omega_B - \omega_D)\right]\langle a^\dagger \sigma_{gD}\rangle}{g_B^2 + \left[\frac{(\gamma_B - \Gamma_D)}{2} + i(\omega_B - \omega_D)\right]\left[\frac{(\kappa - \Gamma_D)}{2} + i(\omega_{pl} - \omega_D)\right]},$$

and

$$\langle a^\dagger \sigma_{gD}\rangle \approx \frac{-ig_D\left[\frac{(\gamma_B + \Gamma_D)}{2} - i(\omega_B - \omega_D)\right]\langle \sigma_{DD}\rangle}{g_B^2 + \left[\frac{(\gamma_B + \Gamma_D)}{2} - i(\omega_B - \omega_D)\right]\left[\frac{(\kappa + \Gamma_D)}{2} - i(\omega_{pl} - \omega_D)\right]},$$

where $\langle \sigma_{DD}\rangle = \langle |e_D\rangle\langle e_D|\rangle$.